\documentstyle[aps,epsf,twocolumn]{revtex}

\begin{document}
\draft
\title{Magnetic Foehn Effect in Nonadiabatic Transition}
\author{Keiji Saito and Seiji Miyashita}
\address{Department of Applied Physics, School of Engineering \\ 
University of Tokyo, Bunkyo-ku, Tokyo 113-8656, Japan}
\date{\today }
\maketitle
\begin{abstract}
The magnetization curves as a response of sweeping magnetic field in
the thermal environment are investigated using 
the quantum master equation. 
In a slow velocity region where the system almost behaves adiabatically,
the magnetic plateau appears which has been 
observed in the recent experiment of V$_{15}$ 
[ Phys. Rev. Lett. (2000)].
We investigate its mechanism and 
propose that this phenomenon is quite universal
in the quasi-adiabatic transition with small inflow of the heat, 
and we call it 'Magnetic Foehn Effect'. 
We observe the crossover between this mechanism and
the Landau-Zener-St\"{u}ckelberg mechanism changing the velocity. 
Some experiment is proposed to 
clarify the inherent mechanism of this effect.
\end{abstract}
\noindent \\
\vspace*{-0.5cm}
\vspace*{-0.8cm}

Properties of quantum dynamics have been studied extensively
for nanoscale magnets and also in microscopic system with nanostructure
\cite{mn12,fe8,nakamura}.
There the nonadiabatic transition plays important roles.
Landau\cite{Landau}, Zener\cite{Zener}, and St\"{u}kelberg\cite{St} (LZS)
derived the well-known nonadiabatic transition probability 
for a two-level system with a sweeping field. 
It depends on the sweeping velocity and energy gap. 
Although the LZS formula is derived in two-level systems,
it can also describe nonadiabatic transitions at 
the avoided crossings in the uniaxial magnets \cite{RMSGG97,miya9596}. 
Therefore it can be widely applied to analyze phenomena related to 
the nonadiabatic transition in various materials \cite{mn12,fe8,nakamura}.
However real experiments are always done in a thermal
environment. Hence the studies on the nonadiabatic transition 
with an environment are quite crucial for further understanding 
of time-dependent phenomena in magnetic systems \cite{KN98}.
There universal aspects of thermal effect independent of 
details of reservoirs are very important.
As an example of such universal aspects, we have studied 
the magnetization process in uniaxial molecule magnets such 
as Mn$_{12}$ and Fe$_{8}$ at very low temperatures 
in thermal environment \cite{SMD00}. There the step-wise magnetization 
process is observed, which is due to the nonadiabatic transition and
a fast damping process to the ground state. 
We call it apparent (or deceptive) nonadiabatic process.  
We also found that the transition probability of purely quantum mechanics
is deducible from this process. This mechanism does not depend on
the detailed structure of reservoir 
apid fluctuation of noise at the resonant field 

In this Letter, we propose another universal qualitative aspect of 
the thermal effect on the quasi-adiabatic transition where the LZS
probability is almost one. We understand that the effect is quite 
general in systems which behave almost adiabatically in the 
dissipative environment.
The present study is directly related to
the recent experiment for the molecule V$_{15}$ which is 
effectively regarded as two level system \cite{V15,CWMBB}.
This molecule has $15$ atoms of V but they divided into subclusters 
of $6$,$3$, and $6$ spins.
The subcluster of $6$ spins forms a singlet state and contributes little
to the magnetization and only the cluster of $3$ spins mainly contributes 
to the magnetization. Therefore the model can be regarded as a two 
level system. This molecule is a very simple system and 
we may expect to see the LZS process clearly \cite{Landau,Zener,St}.
However it was observed that the scattered population (i.e., that 
at the excited level) decreases when the sweeping velocity becomes faster 
\cite{CWMBB}.
This is opposite to what we expect in view 
point of the LZS mechanism where the probability of the adiabatic 
transition should decrease for faster change of the field and the 
population of the excited state should increase. 

Chiorescu et al. \cite{CWMBB} explained this behavior in the view
of the phonon bottleneck effect which means 
a lack of phonon number which contributes to the excitation from the 
ground state at near resonant magnetic fields.
In the experiment, the heat reserver has a double-structure, i.e.,
the spin system is attached to the phonon system of the crystal, and
the phonon system is attached to the external reservoir which is the liquid 
He.
The contact between the phonon system and the external reservoir 
is so week that in short time scale, thermal effect in
the spin system is caused by only phonons.
Due to this effect, the population of the excited state 
does not increase enough, and saturates at some value,
which causes a magnetic plateau.

In this Letter, we show that plateau in the magnetization curve is 
also observed in the case that the spin system is connected with 
a single heat reserver with slow relaxation rate, 
and propose that this qualitative property is 
universal with regardless to the detailed structure of environment.
We investigate a magnetization process for a sweeping field
by making use of the quantum master equation
which we have used in our studies of quantum dynamics in dissipative 
environments \cite{SMD00,MSK00}. Thereby, we investigate the
magnetization plateau for various sweeping velocities and temperatures.

The Hamiltonian we shall consider is given by 
\begin{equation}
{\cal H} = H(t) \, S^{z} + \Gamma \, S^{x},
\label{hamil}
\end{equation}
where $H(t)$ is the sweeping field, $H (t) = vt $
and $\Gamma$ is the transverse field. 
Transverse field represents a term causing quantum fluctuation and  
does not commute with the magnetization $S_z$.
This simple system is realized in many cases, e.g.,
the isotropic anti-ferromagnetic Heisenberg chain 
with odd number of spins has the doublet in the ground state.
Actually V$_{15}$ is in this situation \cite{CWMBB}.

For this system (\ref{hamil}), the LZS transition probability is given by, 
\begin{equation}
P_{\rm LZS}(v) = 1 -\exp\left[ -\frac{ \pi \Gamma^2 }{ v } \right],
\label{LZSpro}
\end{equation}
in the case of $S=1/2$ spin system \cite{Landau,Zener,St}. 
Thus the normalized magnetization
at $t=\infty$ is given by,
\begin{eqnarray}
M_{\rm out} = 1- 2 P_{\rm LZS}(v).
\label{lzsmag}
\end{eqnarray}
We should note that this expression of $M_{\rm out}$ is also exact
for any values for $S$ although (\ref{LZSpro}) is derived for $S=\frac{1}{2}$.

We introduce a thermal environment taking the phonon system as the bath, 
${\cal H}_{\rm B}= \sum_{\omega} \omega b_{\omega}^{\dagger}b_{\omega} $,
where $b_{\omega}$ and $b_{\omega}^{\dagger}$ are the 
annihilation and creation boson operators of the frequency $\omega$.
We adopt the spectral density of the boson bath $I(\omega )$ in the form
$I (\omega ) = I_{0} \omega^{\alpha}\, (\omega \ge 0),\,\, 0\, (\omega < 0)$, 
with $\alpha=2$.
In the experimental situation in the magnetic molecules
such as $S=10$ in Mn$_{12}$ and Fe$_8$,
the hyperfine interaction and the dipole interaction are not negligible 
at very low temperatures \cite{PS96,CFRAV99}. 
In the case of V$_{15}$, the phonon gives a dominant contribution \cite{CWMBB}.

In the case of phonon bath, we can derive an 
equation of motion of the reduced density matrix $\rho$ tracing 
out the degree of freedom of the bath
in the following form (the quantum master equation \cite{QME}):
\begin{equation}
\frac{\partial\rho(t)}{\partial t} = 
\frac{1}{i\hbar}\left[ {\cal H}, \rho (t) \right] -\lambda
\left( \left[X,R\rho(t)\right] + \left[X,R\rho(t)\right]^{\dag} \right) , 
\label{QMEq}
\end{equation}
where $X$ is a system operator through which the system and the 
bath couple with the constant $\lambda$. The first term of the 
right-hand side describes the pure quantum dynamics of the system 
while the second term represents effects of environments at a 
temperature $T(=\beta^{-1})$. There $R$ is defined as follows: 
\begin{eqnarray}
\langle k | R  | m \rangle &=& 
\zeta (E_{k} - E_{m})
n_{\beta} ( E_{k} - E_{m} )  
\langle k | X  | m \rangle , \nonumber \\
\zeta (\omega ) &=& I(\omega ) -I(- \omega) , 
\quad {\rm and} \quad
n_{\beta}( \omega ) =  ({e^{\beta\omega} -1 })^{-1}, \nonumber 
\end{eqnarray}
where $| k \rangle $ and $| m \rangle $ represent the eigenstates 
of ${\cal H}$ with the eigenenergies $E_{k}$ and $E_{m}$, respectively.

We simulate the evolution given by Eq. (\ref{QMEq}) for various sweeping 
velocities in $1/2$ spin system of (\ref{hamil}). 
Throughout this Letter, we set $\hbar $ to be unity.
From now on, we set parameters as $\Gamma = 0.5, T =1.0$, 
and $\lambda=0.001$. In Fig.1(a), we present the magnetization 
curves for fast sweeping rates, $v=0.1, 0.2$, and $0.4$.
Here we clearly find that the magnetic plateau decreases when $v$ 
increases, which is consistent with (\ref{lzsmag}).
On the other hand, in the case of slow sweeping rates 
we also find the magnetic plateau as shown in Fig.$1$(b)
although in these sweeping rates the LZS transition probabilities 
(\ref{LZSpro}) are almost one. Here we should note that the 
magnetic plateau increases when when $v$ increases, which is 
an opposite property to the fast sweeping case. This is the same 
phenomenon as the experiment \cite{CWMBB}. 
\vspace*{-0.7cm}
\begin{figure}
\noindent
\centering
\epsfxsize=8.5cm \epsfysize=6.5cm \epsfbox{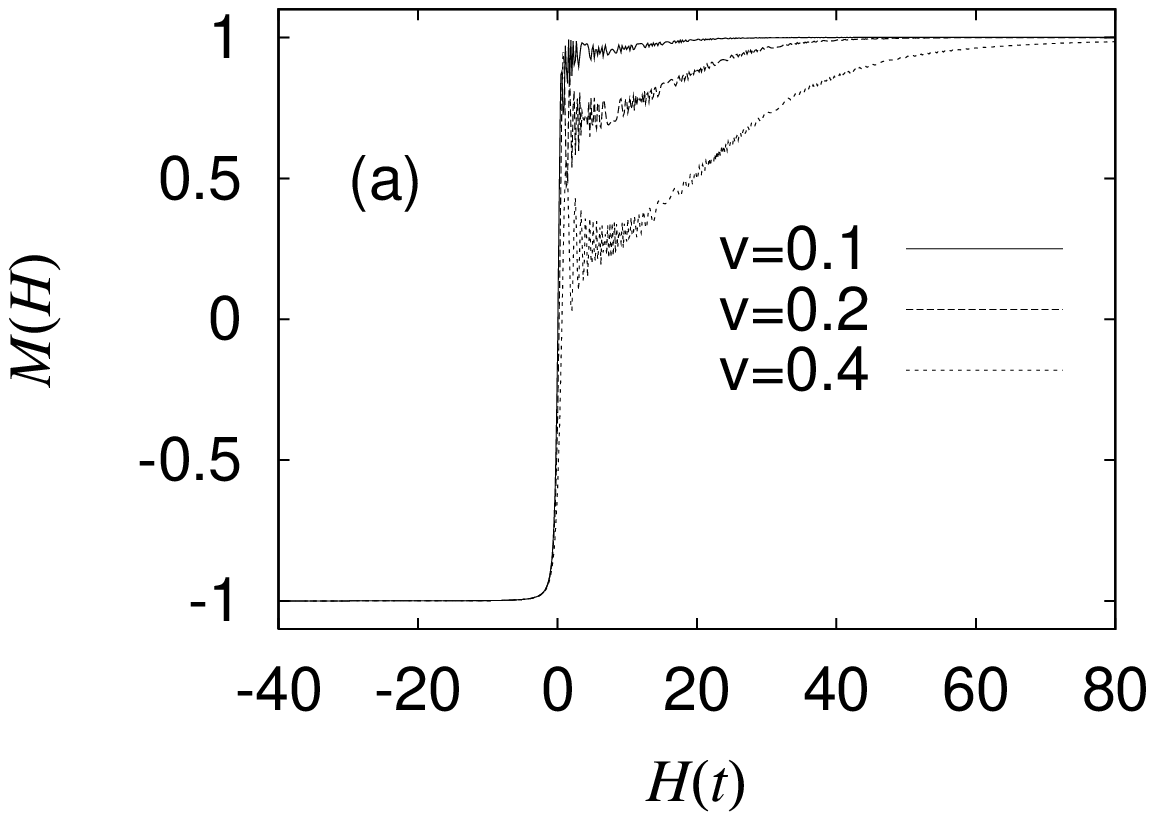} 
\\
\noindent
\centering
\epsfxsize=8.5cm \epsfysize=6.5cm \epsfbox{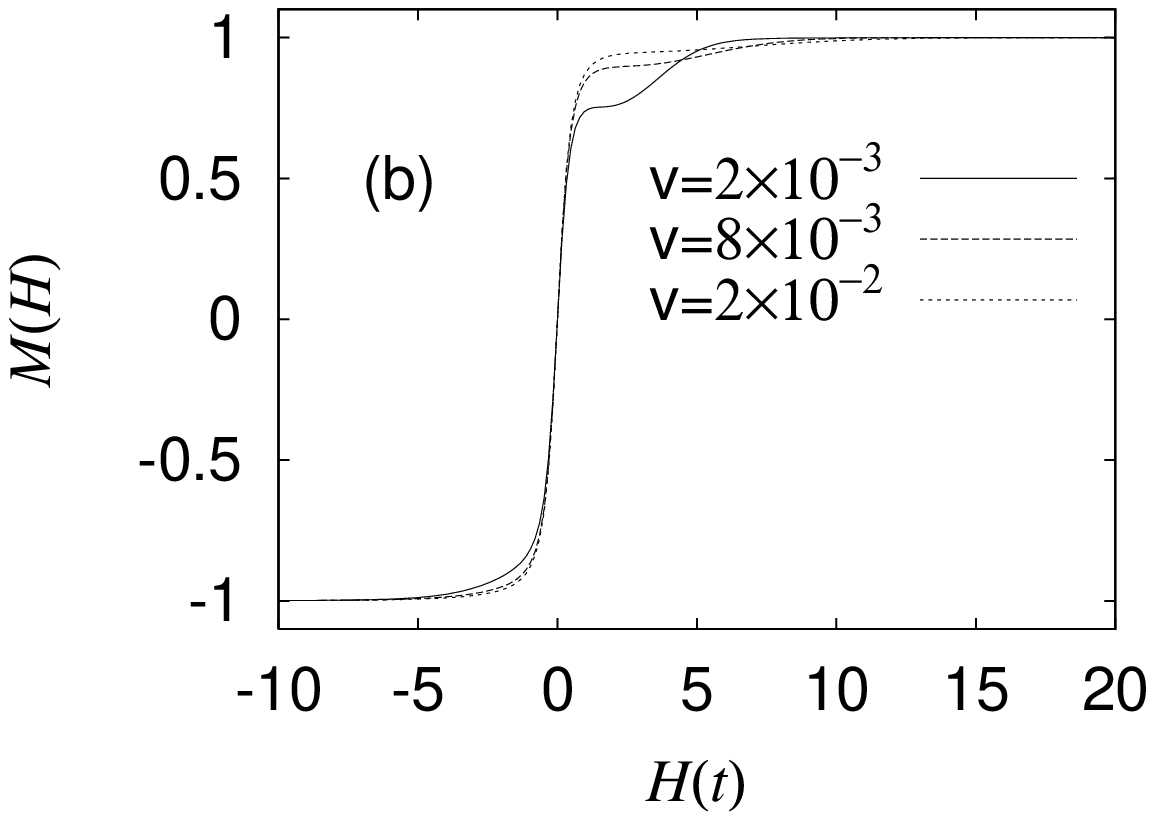}
\caption{Magnetization process for (a)$v=0.1, 0.2$, and $v=0.4$, and 
(b) $v=0.001, v=0.006$, and $v=0.01$}
\end{figure}

In Fig.2 we show the nonmonotonic dependence of the plateau height 
$M_{\rm out}$ on velocities. Trivially when $v$ is very large,
$M_{\rm out}$ shows linear dependence on $1/v$ from Eq. (\ref{lzsmag}). 
In the slow sweeping rate region with $P_{\rm LZS} \sim 1$, 
$M_{\rm out}$ goes down.

\vspace*{-0.7cm}
\begin{figure}
\noindent
\centering
\epsfxsize=8.5cm \epsfysize=6.0cm 
\epsfbox{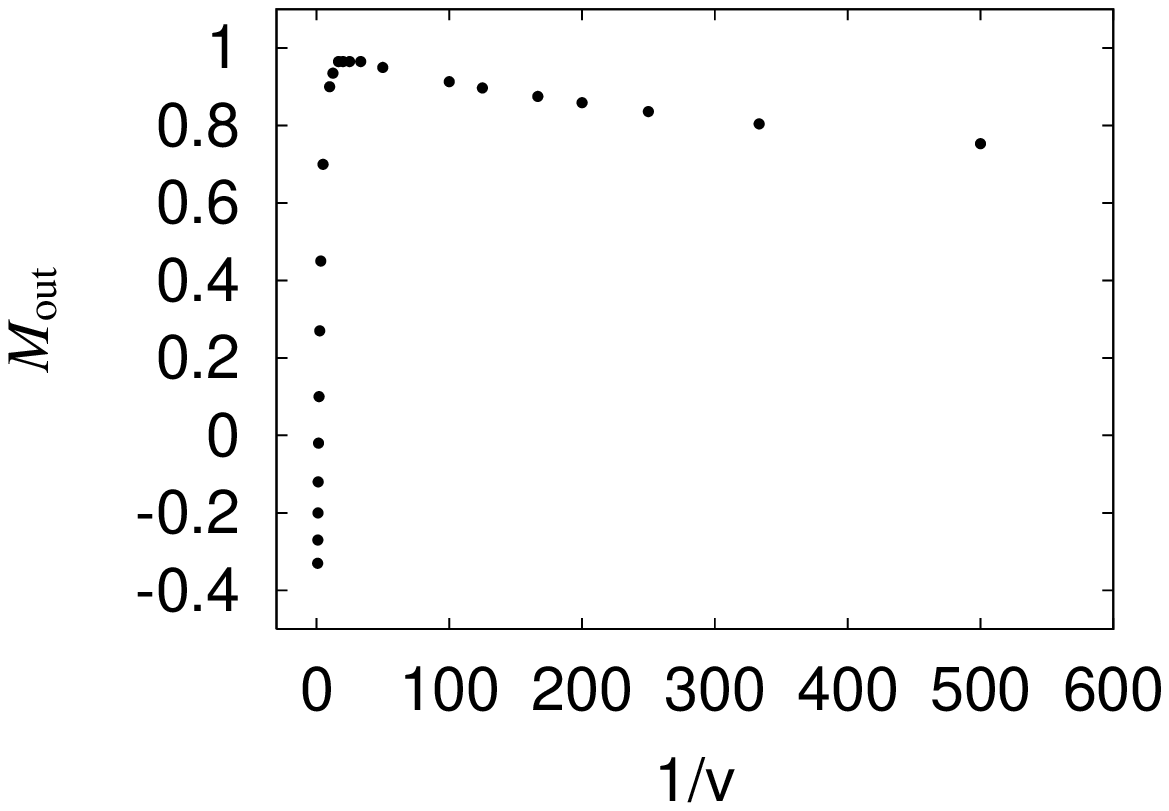} 
\caption{The plateau height, $M_{\rm out}(v)$, 
as a function of the sweeping rate}
\end{figure}

Here we discuss the population in the excited level $\rho_{22}(H)$ 
for various sweeping velocities.
Time dependences of $\rho_{22}(H)$ are shown in Fig.$3$, where 
the population of $\rho_{22}(H)$ changes as a function of
the sweeping velocity.
Because the matrix element of the operator $R$ is proportional 
to $\lambda$ and $\Delta^{\alpha},(\Delta = \sqrt{ H^2 + \Gamma^2})$ , 
the thermalization rate $\gamma (\Delta , \lambda)$ is proportional to 
these values as $\gamma (\Delta , \lambda ) \propto \lambda \Delta^{\alpha} $.

Let us investigate the relation between the thermalization rate 
$\gamma (\Delta , \lambda)$ and the sweeping velocity $v$.
If $v$ is much larger, i.e., $v \gg \gamma (H, \lambda )$, 
then no thermal relaxation occurs. 
In this case, abrupt change in the distribution $\rho_{22}$ due to the 
nonadiabatic transition takes place at $H=0$. This behavior
is demonstrated in the case of $v=2\times 10^{-1}$, and 
$4\times 10^{-1}$ in Fig.$3$. This sudden change causes 
the magnetic jump as shown in Fig.1(a).
Here it should be noted that we see the precession which means that 
the state is still highly coherent. 
In this region of the velocity, when the velocity becomes 
small, the plateau goes up because $P_{\rm LZS}$ monotonically increases 
until a simple adiabatic magnetization curve appears due 
to $P_{\rm LZS} \sim 1$.

In the further slow velocity, thermalization process begins to take place,
that is, $\rho_{22} (H)$ tends to relax to its equilibrium value,
\begin{eqnarray} 
\rho_{22}^{\rm eq}(H)={e^{-\beta\Delta }\over 1+e^{-\beta\Delta } }.
\end{eqnarray}
This increase of $\rho_{22}(H)$ causes a magnetic plateau 
as shown for $v=2 \times 10^{-3}$, and $8 \times 10^{-3}$ in Fig.$3$. 
In order to realize a visible plateau, thermalization rate should be small
comparing with $v$. 
Actually, because of $\gamma (H, \lambda ) \propto \lambda \Delta^{2} $,
thermalization is very slow at around the resonant point due to 
small $\Delta $.  
We also studied the system with $\alpha=0$, where we found that the
plateau sustain for large values of $H(t)$ and the shape of the magnetization
process seems very different.

\begin{figure}
\noindent
\centering
\epsfxsize=8.5cm \epsfysize=6.5cm \epsfbox{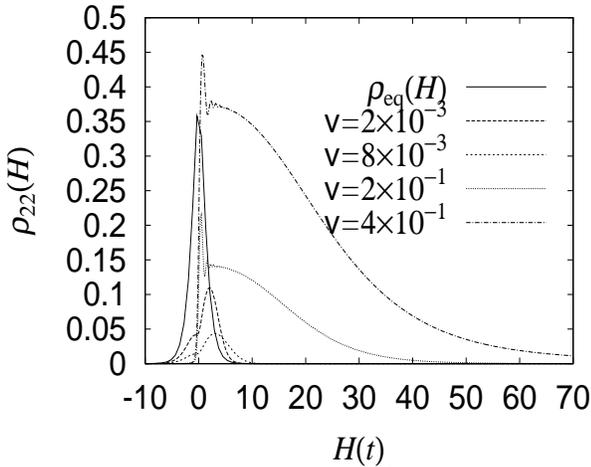} 
\caption{Field dependence of  $\rho_{22}$: (a) the equilibrium value
and (b) $\rho_{22}$ in the sweeping field}
\end{figure}

We associate the mechanism of magnetic plateau 
with the well-known Foehn phenomenon in the meteorology. 
The air with the vapor climbs up the mountain getting 
colder adiabatically and next the rain ensues. At this moment the vapor 
gives the heat to the air as the latent heat.
Then, the air alone goes down the mountain and
the temperature of the air 
increases higher than the original one due to the inflow of the latent 
heat at the mountain. 
 In the present magnetic system, 'climbing up the mountain' corresponds 
to 'sweeping of magnetic field to $H=0$', 'the latent heat' to 
'the heat from the phonon', and 'the increase of temperature of the air' to 
'the increase of temperature of the spin'. 
After the plateau the magnetization relaxes to the equilibrium value due to 
cooling by the thermal bath, which corresponds to 
the cooling of the air by the land after hot air reaches to the ground.
Thus these similarities lead us to call this magnetic phenomenon 
'Magnetic Foehn effect'. We confirmed that the Magnetic Foehn effect is 
also observed in the Hamiltonian (\ref{hamil}) with larger $S$. 
The essential mechanism of this phenomenon is inflow of heat during 
adiabatic process. Because the mechanism is quite simple,
we can say that the 'Magnetic Foehn effect' is a universal 
phenomenon in the magnetic systems which behave almost adiabatically.

Because the LZS transition occurs
only in the vicinity of $H=0$ and at other points only relaxations due to the
dissipation term occur,
the time evolution process may be divided into three regions:$
\rho (t) \sim {\cal U}_{\rm d}(t:0) \cdot  {\cal U}_{\rm LZS} \cdot
{\cal U}_{\rm d}(0:-\infty ) \rho (-\infty )$,
where ${\cal U}_{\rm d} (t:s) $ is the time evolution operator due to 
dissipation part from a time $s$ to $t$, and 
${\cal U}_{\rm LZS}$ represents the 
LZS scattering matrix which corresponds to the pure quantum 
dynamics in the first term in (\ref{QMEq}) \cite{MSD98}. 
For slow velocity case $v \ll 1$, we can consider that the system 
behaves adiabatically and thus we put the 
evolution ${\cal U}_{\rm LZS}$ to be unity.
Thus the transfer between the levels occurs only by the dissipation 
${\cal U}_{\rm d} (t:s) $.
If we write down the second term of (\ref{QMEq}) explicitly,
we have the following equations for the diagonal term
\begin{eqnarray}
\dot{ \rho}_{22}  = -\frac{1}{2}X_{12} 
\lambda \, n_{\beta} \left( \Delta \right)  \,
\zeta \left( \Delta \right) \, 
\left[  ( e^{\beta \Delta } +1 )\rho_{22} - 1 \right] ,
\label{AE}
\end{eqnarray}
where $X_{12}$ is the matrix elements of the operator $X$.
In our simulation, it is expressed as $X_{12} = \frac{\left| 
H -\Gamma \right| }{\sqrt{H^2 + \Gamma^2}}$ for $X=S_x+S_z$ .
The term $X_{12}$ depends on the choice of $X$, and is not universal.
This equation (\ref{AE}) leads the Magnetic Foehn effect.

At the end of this Letter, let us discuss some experimental situations.
As for the observation of the crossover from the Magnetic Foehn region to 
that of LZS region is difficult in V$_{15}$, because 
the crossover from the Magnetic Foehn region to the LZS region 
locates at a very fast sweeping rate. 
For example, if $\Gamma=0.1$K, the crossover sweeping rate $v$ is of 
order $10^7$ H$/$sec. Such fast change of the field is not easy to realize.
However when the gap become small, the field at the crossover sweeping rate $v$
becomes small.  In Fig.$4$, 
we show the dependence of the gap on the size and anisotropy for the system:
\begin{equation}
{\cal H}=J\sum_{i} \left( S_i^xS_{i+1}^x+S_i^yS_{i+1}^y+AS_i^zS_{i+1}^z 
\right) ,
\end{equation}
where ${\bf S}_i$ is an $S=1/2$ spin and the open boundary condition is
adopted. The gap becomes small with $A$, and decreases with $N$ 
exponentially.
Thus we expect that the crossover 
would be observed in some system and there we can obtain 
many informations of effects due to combination of the quantum 
process and the thermal effect.
\begin{figure}
\noindent
\centering
\epsfxsize=8.5cm \epsfysize=6.0cm 
\epsfbox{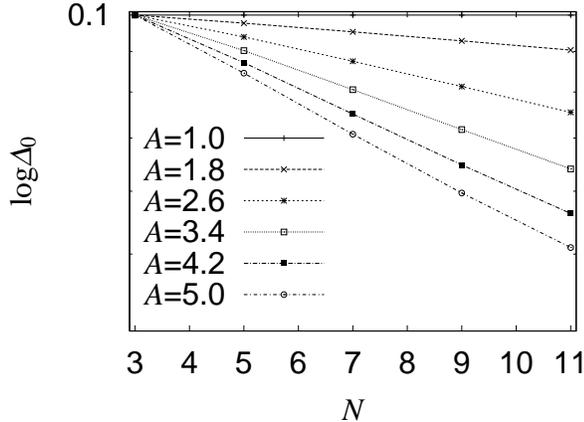} 
\caption{Dependence of the gap on the size of the system $N$ and 
the anisotropy $A$}
\end{figure}

In V$_{15}$, Chiorescu et al. attributes the Magnetic Foehn phenomenon
to the lack of phonon as a mechanism of the insufficient supply of heat. 
In such a situation, we would 
propose an experiment to realize the adiabatic transition
of an isolated spin system. That is,
we first sweep the field from the large negative value to the field 
($H_{\rm p}$) 
where the magnetic plateau $(M_{\rm p})$ is observed.
At this point the number of phonon is very small and supply of phonons from 
out side is expected to be slow.
In this circumstance, if the field is swept in the opposite 
direction from $H_{\rm p}$ to 
$-H_{\rm p}$,
the spins at the lower spin can not be excited by the phonon because 
no phonon is available, and behaves pure quantum mechanically.
In the experiment the LZS probability is almost one. 
The spins at the higher level may emit the phonon and relax to the
lower level. But the emitted phonon will be used to excite the spin
again because the population of the upper level is much smaller than that of
equilibrium. 
Thus we expect that the magnetization simply changes the sign when
$H_{\rm p} \rightarrow -H_{\rm p}$.
 In the iteration of this process $H_{\rm p} \rightarrow -H_{\rm p}
 \rightarrow H_{\rm p} \rightarrow -H_{\rm p}  \rightarrow \cdots$, the 
magnetization would maintain the same amplitude 
for a while, $-1 \rightarrow 
M_{\rm p} \rightarrow -M_{\rm p} \rightarrow M_{\rm p}
\rightarrow -M_{\rm p}\cdots$, 
before the heat flows in from the external environment 
and equilibrates the system. From this slow relaxation of magnetization
we could know the relaxation rate between the phonon and the external bath.
On the other hand, in the Magnetic Foehn phenomenon due to slow relaxation 
but not short of phonon number, 
$M_{\rm p}$ relaxes with the thermalization rate. From Eq.(\ref{AE}), 
we can derive the relation of $\rho_{22}$ for the iteration:
$\rho_{22}^{(n)} = p_1 + p_2 \rho_{22}^{(n-1)}$ with 
$\rho_{22}^{(0)} = 0$ and $p_1 = (1- M_{\rm p})/2$.
The sequence $M_{\rm p}^{(n)}$ is given by 
$|M_{\rm }^{(n)}| = 1- 2\rho_{22}^{(n)}$
where $\rho_{22}^{(n)} = (1-M_{\rm p}) (1-p_{2}^{n} )/(1-p_2 ) $. Here 
$p_{2}$ is given by 
\begin{eqnarray}
p_{2} = \exp \left( -
\lambda ' \int_{-H_{\rm p}/v}^{H_{\rm p}/v }  \, d\tau  \,
\Delta^{\alpha }(\tau ) \coth ( \beta \Delta (\tau )/2 ) \right).
\end{eqnarray}

We would like to thank Professor B. Barbara and Dr. I. Chiorescu 
for valuable communications for their work of the reference\cite{CWMBB}.
The present work is partially supported by Grant-in-Aid for Scientific 
Research from Ministry of Education, Science, Sports and Culture of Japan.


\begin{thebibliography}{000}
\bibitem{mn12}
L. Thomas, F. Lionti, R. Ballou, D. Gatteschi,
R. Sessoli and B. Barbara, Nature {\bf 383}, (1996) 145.
,L. Thomas et al., Nature {\bf 383}, (1996) 145.
,F. Lionti, L. Thomas, R. Ballou, Barbara, A. Sulpice,
R. Sessoli and D. Gatteschi, J. Appl. Phys. {\bf 81}, (1997) 4608.
\bibitem{fe8}
W. Wernsdorfer, et al. Phys. Rev. Lett. {\bf 78} (1997) 1791; 
{\bf 79} (1997) 4014.
\bibitem{nakamura}
H. Nakamura: {\em Dynamics of Molecules and Chemical Reactions} (Marcel Dekker, 1996) 473
\bibitem{Landau} L. Landau, Phys.\ Z. Sowjetunion {\bf 2}, 46 (1932).
\bibitem{Zener} C. Zener, Proc.\ R. Soc.\ London, Ser.\ A {\bf 137}, 
696 (1932).
\bibitem{St} E. C. G. St\"uckelberg, Helv.\ Phys.\ Acta {\bf 5}, 369 (1932).
\bibitem{RMSGG97}
H. De Raedt, S. Miyashita, K. Saito, D. Garc\'{i}a-Pablos, and N. Garc\'{i}a,
Phys. Rev. B, {\bf 56} 11761 (1997).
\bibitem{miya9596} S. Miyashita, J. Phys. Soc. Jpn. {\bf 64}, (1995) 3207;
{\bf 65}, (1996) 2734.
\bibitem{KN98} Y. Kayanuma and H. Nakamura, Phys. Rev. B{\bf 57} (1998) 13099.
\bibitem{SMD00}
K. Saito, S. Miyashita, and H. De Raedt, Phys. Rev. B {\bf 60} 14553 (1999)
\bibitem{V15}A. M\"{u}ller, J. D\"{o}ring, Angew. Chem. Int. Ed. Engl., 
{\bf 27}, 1721 (1991). D. Gatteschi, L. Pardi, A. L. Barra, A. M\"{u}ller, 
J. D\"{o}ring, Nature, {\bf 354}, 465 (1991), W. Wernsdorfer, E. Bonet Orozco,
K. Hasselbach, A. Benoit, D. Mailly, O. Kubo, H. Nakano, and B. Barbara, 
Phys. Rev. Lett., {\bf 79} 4014 (1997).
\bibitem{CWMBB}
I. Chiorescu, W. Wernsdorfer, A. M\"{u}ller, H. B\"{o}gge, and B. Barbara,
Phys. Rev. Lett. (2000); cond-mat/9911180/.
\bibitem{MSK00} 
S. Miyashita, K. Saito, and H. Kobayashi, unpublished, cond-mat/9911148/.
\bibitem{PS96}
N. V. Prokof\"{e}v and P. C. E. Stamp, J. L. Temp. Phys. {\bf 104} 143 (1996).
\bibitem{CFRAV99}
A. Cuccoli, A. Fort, A. Rettori, E. Adam, and J. Villain, EPJ (1999), 
cond-mat/9905273/.
\bibitem{QME}
K. Saito, S. Takesue, and S. Miyashita, Phys. Rev. E. {\bf 61} 2397 (2000),
R. Kubo, M. Toda, and N. Hashitsume, 
{\em Statistical Physics II} (Springer-Verlag, New York, 1985).
\bibitem{MSD98}
S. Miyashita, K. Saito, and H. De Raedt, Phys. Rev. Lett. {\bf 80},
1525 (1998).
\end{thebibliography}
\end{document}